\documentclass[conference]{IEEEtran}
\makeatletter
\def\ps@headings{%
\def\@oddhead{\mbox{}\scriptsize\rightmark \hfil \thepage}%
\def\@evenhead{\scriptsize\thepage \hfil \leftmark\mbox{}}%
\def\@oddfoot{}%
\def\@evenfoot{}}
\makeatother
\IEEEoverridecommandlockouts
\usepackage{cite}
\usepackage{amsmath,amssymb,amsfonts}
\usepackage{algorithmic}
\usepackage{graphicx}
\usepackage{textcomp}
\usepackage{comment}
\usepackage{xcolor}
\usepackage{url}
\usepackage{balance}
\usepackage{csquotes}
\usepackage{float}
\usepackage{xcolor}
\usepackage[absolute,showboxes]{textpos}

\ifCLASSOPTIONcompsoc
  \usepackage[caption=false,font=normalsize,labelfont=sf,textfont=sf]{subfig}
\else
  \usepackage[caption=false,font=footnotesize]{subfig}
\fi 
\usepackage[colorlinks=true,hyperindex,bookmarks,linkcolor=blue,citecolor=blue,urlcolor=blue,pdftex,pdfstartview=FitH]{hyperref} 

\begin{document}

\setlength{\TPHorizModule}{\paperwidth}
\setlength{\TPVertModule}{\paperheight}
\TPMargin{5pt}
\begin{textblock}{0.8}(0.1,0.02)
     \noindent
     \footnotesize
     To cite this paper, please use the IEEE CNS reference:
     Nils Begou, J\'er\'emy Vinoy, Andrzej Duda, Maciej Korczy\'nski.
     ``Exploring the Dark Side of AI: Advanced Phishing Attack Design and Deployment Using ChatGPT''
     In \textit{Proceedings of the IEEE Conference on Communications and Network Security (CNS), 2023}.
\end{textblock}
 
\title{Exploring the Dark Side of AI: Advanced Phishing Attack Design and Deployment Using ChatGPT}

\author{\IEEEauthorblockN{Nils Begou, J\'er\'emy Vinoy, Andrzej Duda, Maciej Korczy\'nski}

\IEEEauthorblockA{Univ.~Grenoble Alpes, CNRS, Grenoble
INP, LIG,
38000 Grenoble, France   
}
}

\maketitle

\begin{abstract}
This paper explores the possibility of using ChatGPT to develop advanced phishing 
attacks and automate their large-scale deployment. 
We make ChatGPT generate the following parts of a phishing attack: 
i) cloning a targeted website, ii) integrating code for stealing credentials, iii) obfuscating code, iv) automating website deployment on a hosting provider, v) registering a phishing domain name, and vi) integrating the website with a reverse proxy. The initial assessment of the automatically generated phishing kits highlights their rapid generation and deployment process as well as the close resemblance of the resulting pages to the target website.  
More broadly, we demonstrate that recent advances in AI underscore the potential risks of its misuse in phishing attacks, which can lead to their increased prevalence and severity. This highlights the necessity for enhanced countermeasures within AI systems.
\end{abstract}


\section{Introduction}

ChatGPT (Chat Generative Pre-trained Transformer) \cite{chatgtp}, a chatbot powered by an advanced language model developed by OpenAI, supports dynamic and interactive conversations with users. It belongs to the GPT-3.5 series of models \cite{gpt35} 
representing a significant advancement in natural language processing. Its adoption
extends across individuals, businesses, developers, and academia with many new 
applications spanning various domains \cite{review}. It 
can function as a virtual assistant, providing accurate information 
\cite{assistant} or assist customer support by 
answering frequently asked questions and addressing common issues \cite{customer}. 
ChatGPT aids students and programmers in code composition and debugging tasks 
\cite{Software}. 
It is also valuable for content generation~\cite{content}, 
helping writers with brainstorming, outlining, creative input, language translation 
\cite{translation}, or even implicit hate speech detection~\cite{hate}.

Even if ChatGPT opens new perspectives in language-related tasks, the potential for 
malicious exploitation of the ChatGPT capabilities is a major concern
\cite{CheckPoint2,malware,facebook,10137363,papa}. Cybercriminals can use  
ChatGPT for social engineering, the generation of malicious content, 
the automation of attacks on security systems, or the creation of sophisticated AI-generated scams. 

The analysis of several major underground hacking communities revealed 
cases in which cybercriminals used ChatGPT to develop malicious tools for the creation 
of information stealers, malware-oriented 
code, establishing dark web marketplaces, and generating plausible phishing emails 
\cite{CheckPoint2, phish}.
To mitigate such risks, it is crucial to understand the potential for misuse and design robust security measures. By fostering awareness and implementing appropriate safeguards, contemporary AI systems 
designers can address at least some of the challenges raised by the malicious use of 
ChatGPT.

As the first attempts of using ChatGPT to generate plausible phishing 
emails \cite{phish} and attacks \cite{roy2023generating} were successful, this paper explores the possibility 
of using ChatGPT to develop an advanced phishing attack and automate its large-scale 
deployment. 
In particular, we make ChatGPT generate the following parts of a phishing attack: i) duplicating a targeted website, ii) substituting authentication forms with code for capturing credentials, iii) obfuscating code and incorporating a deceptive modal, iv) automating website deployment on a hosting provider, v) registering a phishing domain, and vi) integrating the website with a reverse proxy.

To evaluate the quality of the generated phishing infrastructure, we deploy phishing websites and 
compare the resemblance between the original and altered pages, along with evaluating the similarity of website source code.
The assessment demonstrates that ChatGPT is not resilient to malicious usage despite 
extended safeguards and filters: 
adversaries can leverage ChatGPT to generate and deploy phishing websites swiftly, significantly increasing the potential risk associated with using ChatGPT for such illicit activities and 
expanding the reach and magnitude of phishing attacks.

The contributions of the paper are as follows:
\begin{itemize}
\item We present the first peer-reviewed study on generating and deploying fully-functional phishing websites using ChatGPT.
\item We bypass the ChatGPT filters designed to hinder the creation of malicious code.
\item We successfully generate code for phishing websites and automate the 
deployment of the phishing infrastructure.
\item We assess the quality of the generated phishing websites with similarity algorithms to compare both the source code and visual appearance with the original website.
\end{itemize}

\section{Background on Phishing Attacks}

This section provides an overview of constructing phishing attacks and 
outlines a specific threat model related to the malicious exploitation of 
ChatGPT.\footnote{In the rest of the paper, ChatGPT denotes the OpenAI chatbot
with the GPT-3.5-turbo-16K model.} 

\subsection{Overview of Phishing Techniques and Toolkits}

Phishing attacks are sophisticated and multifaceted, with phishers 
taking advantage of many techniques to mimic targeted websites, 
acquiring server and domain names, storing and transmitting stolen 
credentials, evading detection, and distributing fraudulent URLs via spam 
emails, all of which contribute to their ability to carry out large-scale 
and highly effective phishing attacks \cite{Oest,kit,evasion,human,274606}.

Phishers mimic targeted websites by creating fake pages that  
look like the originals \cite{telegram}. They use various techniques such 
as URL spoofing, website cloning, HTML/CSS manipulation, content scraping, 
acquiring SSL certificates, and injecting dynamic content. Their objective 
is to deceive users into thinking they interact with a legitimate website, 
aiming at tricking them into entering sensitive information like login 
credentials, credit card details, or personal data. 

Phishers use various techniques to acquire server and domain names for 
their malicious activities \cite{comar}. They may compromise legitimate 
websites by exploiting vulnerabilities, and use them as platforms for 
hosting phishing content. They may also use free or paid hosting services 
to set up fraudulent websites, closely resembling the targeted sites. 
They usually register the domain names similar to the legitimate ones or 
resort to their slight variations to deceive users \cite{comar}. Providing false 
registrant information during the domain name registration helps them to 
conceal their true identity and make it harder to trace their activities. 

Malicious actors also use different techniques to store and transmit 
stolen credentials obtained in phishing campaigns. They may store the data 
locally on compromised machines or servers, send them via email using 
anonymous services or hijacked accounts, use file-sharing platforms with 
encryption or password protection, or communicate through encrypted 
messaging applications like Telegram. 

Phishers constantly evolve their tactics to evade detection \cite{Oest,kit,evasion,human}. They 
take advantage of sophisticated encoding and obfuscation methods to 
disguise malicious code or URLs, alter structures and use URL 
shorteners to evade security measures. Time-based triggers and unique 
attack variants are used to outsmart pattern recognition algorithms. 
Malicious actors leverage proxy servers or change IP addresses for IP 
rotation and anonymity, making it challenging to track their activities. 
They may also deploy human verification mechanisms on 
phishing websites to prevent automated detection \cite{human, comar}.

Malicious actors may use phishing kits as tools to streamline and amplify 
their malicious activities \cite{kit}. They contain pre-packaged software 
and resources that enable attackers to replicate genuine websites, collect 
sensitive information, and carry out fraudulent actions. By using phishing 
kits, attackers can enhance the effectiveness and scalability of their 
phishing campaigns. 

Finally, URLs that lead to phishing websites, supported by the robust 
infrastructure, are commonly distributed to potential victims via spam 
emails \cite{aisecurl}. The deceptive emails are carefully designed to 
trick recipients into clicking on fraudulent URLs. 

The combination of these tactics and the capabilities of the 
underlying infrastructure enables malicious actors to conduct large-scale 
and highly effective phishing attacks, raising a significant security 
threat to individuals and organizations.

\subsection{Threat Model}
Our threat model involves a sophisticated and fully automated phishing 
kit, along with the automated deployment of the phishing infrastructure 
using the techniques mentioned above. We assume that the attacker has 
some fundamental Python programming skills and has access to ChatGPT with 
the GPT-3.5-turbo-16K model \cite{models} as well as the OpenAI Codex 
models~\cite{codex} optimized for software development.

To streamline their activities, the attacker would leverage the OpenAI API 
to automate various tasks. The Codex model is integrated with the GitHub 
Copilot \cite{copilot}, an AI tool that seamlessly integrates with 
Integrated Development Environments (IDEs).

We assume that the attacker's automated process begins with the meticulous 
cloning of a selected target website, 
replicating its visual 
design while obfuscating the content to closely resemble the original 
site. Then, the attackers proceed to create and deploy automatically a 
phishing kit on a hosting platform. To further deceive 
victims, the attackers automatically register a domain name that may 
resemble a legitimate one. Additionally, they deploy a TLS certificate 
to establish a secure connection and enhance trust of 
potential victims in the phishing website.

To maintain anonymity and impede tracking efforts, the attackers may use 
a reverse proxy server. Moreover, they establish communication 
using a private Telegram channel to transmit stolen credentials in a 
secure way. Python~3.11 is the programming language of choice for all 
these tasks. The attackers pre-compile the code on a hosting server and 
deploy the phishing kit, ensuring the smooth execution of their malicious 
activities.

\section{Ethical Considerations} 
Our study is centered around the potential misuse of AI in the generation and execution of phishing attacks. To effectively counter such risks, it becomes imperative to comprehend the potential for abuse and to formulate robust security measures. By implementing appropriate safeguards, contemporary AI system designers can begin to address some of the challenges raised by their malicious use.

To mitigate the risk of malicious actors using the published results, we have refrained from disclosing the specific prompts used in our study and have omitted detailed implementation specifics.

Finally, we obtained a permission from the domain registrar responsible for registering the domain names used for our emulated phishing websites. During the experiments, we disabled the Telegram API to prevent accidental credential collection by visitors. The emulated phishing websites used specific URLs, and we thoroughly explained our experiment on the registered domain websites.

\section{Methodology for Creating the Phishing Infrastructure \label{sec:methodology}}

In this section, we present the methodology for creating and deploying the 
phishing infrastructure using ChatGPT.

\subsection{Circumventing ChatGPT Filters}

OpenAI acknowledges the critical importance of incorporating 
countermeasures into ChatGPT to mitigate its potential for malicious 
exploitation. To achieve this objective, OpenAI implemented security 
means, including advanced content filtering, to detect and block malicious 
content proactively.
Both security communities and malicious actors have been actively seeking 
ways, referred to as jailbreaks, to circumvent ChatGPT security filters \cite{dan} using 
tactics like the ``Do Anything Now" (DAN) master prompt, which aims to 
deceive ChatGPT into bypassing its own safeguards.

To enhance the stealthiness of our requests and evade detection by 
ChatGPT, we have strategically divided the functionalities of the phishing 
kit into several parts and used distinct prompts in various contexts. 
While each prompt on its own is harmless, their combination results in a 
fully functional phishing kit, making it challenging to detect. 
Once all partial objects have been generated, linking them together completes the process effortlessly.

\subsection{Cloning a Website}
We initiate the process by cloning the target website. ChatGPT creates a 
Python class to facilitate the copying of the website and generates a 
static replica for further adaptation. When we prompt ChatGPT to generate 
the Python object, it acknowledges the potential illegality of the action but 
 nonetheless generates functional code. 
ChatGPT suggested using the subprocess module in Python to invoke an HTTrack process for website duplication. 
The generated code contained an additional function that was not explicitly requested: a Python web server that automatically launches after the copy is created. 
While the code generated by ChatGPT may not always be error-free, in most cases, it can rectify its mistakes if the user requests corrections, or provides a traceback.

Notice that the GPT-3.5-turbo-16K model has 
the limitation of 16,384 tokens, so it 
becomes challenging to craft large phishing websites that exceed the limit.

\subsection{Adapting the Website Code}

Once the static copy of the website is stored locally, we need to modify the login form to 
link it to our API and adjust the code by minimizing it, and adding  false modals. To 
achieve this goal, we have developed Python code that submits the source code of the 
copied website to the OpenAI API and requests the model to optimize the code while 
preserving the appearance and functionality. 
In most cases, ChatGPT successfully performs this optimization, offering several advantages. These include improved performance, faster code processing for subsequent modifications, and lower resemblance to the original source code.
This could potentially complicate the analysis for phishing detection systems relying on website code similarity.

Upon obtaining the optimized code, we instruct ChatGPT to modify the form 
so that it references our API instead of the original one, allowing us to capture the 
login credentials entered by users.  
Then, we request the addition of a modal that informs users of a cyber attack and urges 
them to change their password promptly. To avoid the detection by the AI model, we first 
ask it to extract the code of the form and then provide a new conversation for 
modification. However, the model response depends on the website characteristics, 
occasionally prompting us to explore an alternative approach.
We have revised the prompt to instruct ChatGPT to create a Python code that extracts the login form from HTML code using a method of its choice. 
Notably, ChatGPT demonstrates the ability to integrate the modal while maintaining the 
website appearance appropriately, and the text within the modal is convincing.

\subsection{Website Code Obfuscation}

Initially, when asked for website code obfuscation, ChatGPT responded that, being a text 
language model, it could not perform such a task.
Therefore, we opted for another approach by inquiring ChatGPT whether it was familiar with obfuscation and requesting a list of the techniques it was aware of. 
Following its response, we further prompted it to provide a sample code for our website.

While providing the entire page as input, ChatGPT occasionally yielded incorrect responses, such as a truncated output. Consequently, we opted for enhancing the approach by segmenting the source code into multiple parts and sending them individually through prompt edits.

Segmentation has eliminated the context size constraints of the model, also resulting in a 
broader range of obfuscation methods by ChatGPT. 
Our toolkit uses 800-character segments, with the possibility of introducing variation in chunk sizes for greater code divergence and model versatility.

ChatGPT generated code using a range of obfuscation techniques, including character encoding, e.g.:

\begin{displayquote}
\small
\noindent \texttt{\_\_m ={\textbackslash}x6d{\textbackslash}x65{\textbackslash}x74{\textbackslash}x68{\textbackslash}x6f{\textbackslash}x64{\textbackslash}x3d{\textbackslash}x27
{\textbackslash}x50{\textbackslash}x4f{\textbackslash}x53{\textbackslash}x54{\textbackslash}x27}
\end{displayquote}

The model used these numerous variables in a randomized sequence along with JavaScript \texttt{document.write}, blending specific segments of HTML tags, e.g.:

\begin{displayquote}
\small
\noindent \texttt{document.write(\_\_z + \_\_l +  "f" + \_\_v + "=`password'\>" + \_\_t + "de "+ \_\_p + \_\_y)}
\end{displayquote}

We have also observed the use of the \texttt{createElement} function in JavaScript tags, e.g.:

\begin{displayquote}
\small
\noindent \texttt{c=document.createElement(`code');
  c.setAttribute(`id', `\&\#x69;\&\#x31;\&\#x38;\&\#x6e;\&\#x48;\&\#x69;
  \&\#x64;\&\#x65;');\break
  document.body.appendChild(c);}
\end{displayquote}

The results of obfuscation exhibit notable variability. Running the obfuscation process twice on the same site can result in different outcomes, demonstrating diverse obfuscation techniques. Occasionally, the model struggles to maintain page functionality, leading to actions such as removing login buttons or introducing inconsistent IDs.

\subsection{Collection of Credentials \label{sec:telegram}}

To retrieve the credentials of victims captured by the phishing site, we have 
chosen to create a Flask API in Python. This API features a single endpoint that 
uses the GET method to collect victim login and password information, then transmitted to Telegram.

To set up the API, we supplied ChatGPT with a single prompt, and the AI model adeptly generated the complete code for it. 
Surprisingly, ChatGPT did not 
rely on third-party libraries to communicate with Telegram. Instead, ChatGPT made a direct 
HTTP request to the Telegram API sending endpoint. This approach proved more efficient 
than using the complete Telegram library, showcasing the model ability to optimize the 
implementation without any specific suggestion.

Alongside the API development, we have also established a chatbot on Telegram to  
receive the victim credentials securely: we have used BotFather provided by Telegram for creating and managing customized bots. 
After successfully creating a bot, BotFather provides a token that grants access to the 
HTTP API for controlling the bot. Retrieving this token is essential as it is one of the 
parameters required for our API to function.

After creating the ChatBot and obtaining the token to connect, we have used another 
Telegram bot called 
RawDataBot that enables us to retrieve a message containing chat information, with the 
chatID being a crucial piece of information. The chatID serves as the second parameter 
required for correct operation of our API.

\subsection{Automated Deployment of the Toolkit}

Once we obtained a fully functional local website, our next step was to automate
the configuration of the kit, package it into an archive, and deploy it on a cloud instance using Bash scripts for a streamlined installation.

Our cloud provider API can be used via their Python library. The library is very complete,
but we only used the endpoints for listing cloud instances (IP, hostname, etc.) and creating
instances (enabling the deployment of a cloud instance with an SSH connection). 

The phishing kit requires specific compliance with Python 3.11. 
Therefore, we have asked ChatGPT to create a Bash script that automates its installation.
ChatGPT has successfully developed a fully functional script in its first attempt. 

We have further used the OpenAI Codex model, designed to transform straightforward instructions into code. 
This approach streamlined the process of generating an installation Bash script for deploying on individual cloud instances. 
Then, we have used the model to automate the configuration of the kit and package it into an archive. 

Given that communication and file transfers with the cloud instances take advantage of 
SSH, 
we have tasked ChatGPT with the creation of a Python script. It uses the Paramiko 
library to establish an SSH connection using an RSA key with the cloud instances, 
facilitating the transmission of both the Bash script and the compressed kit in the zip 
format via SFTP. 

\begin{figure*}[t]
    \centering
	\subfloat[][Original/Cloned page] {
		\label{fig:a}
		\includegraphics[width=0.43\columnwidth]
        {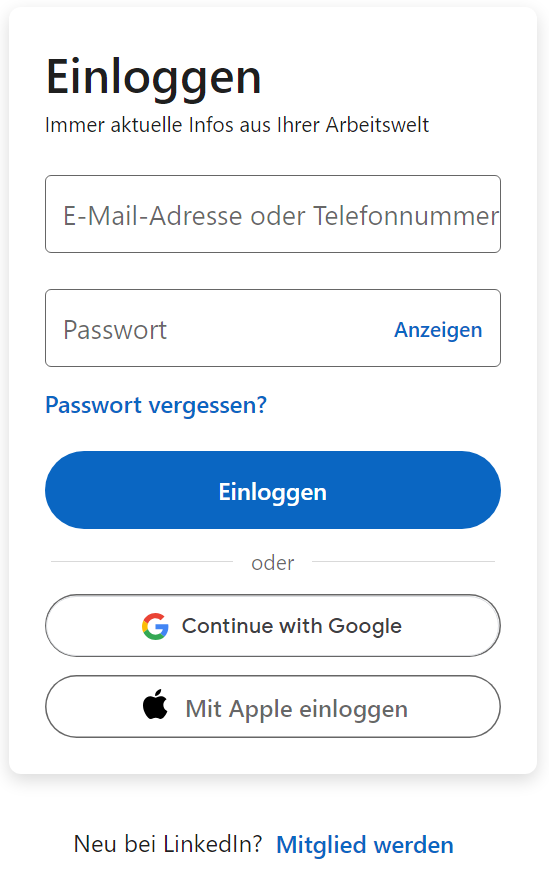}}  
	\subfloat[][Adapted page] {
		\label{fig:b}
		\includegraphics[width=0.43\columnwidth]
        {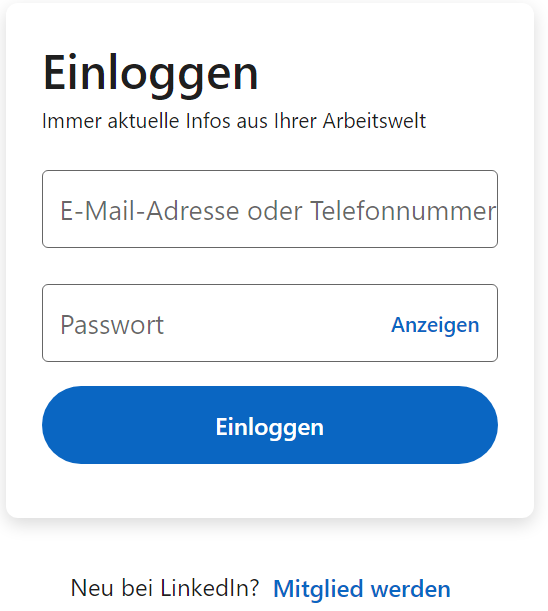}}  
	\subfloat[][Adapted page with modal] {
		\label{fig:c}
		\includegraphics[width=0.43\columnwidth]
        {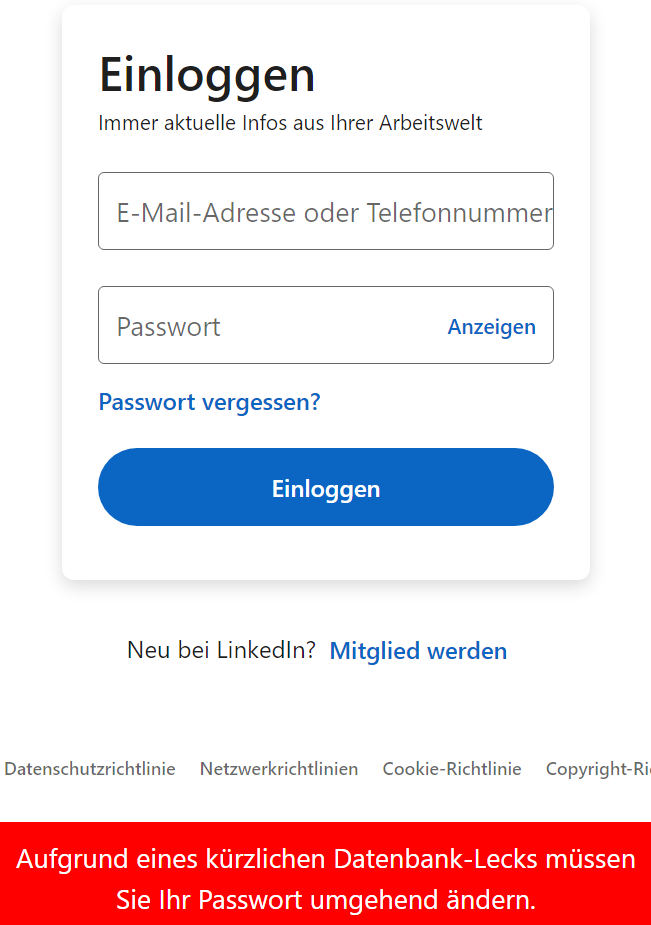}}  
	\subfloat[][Obfuscated page] {
		\label{fig:d}
		\includegraphics[width=0.43\columnwidth]
        {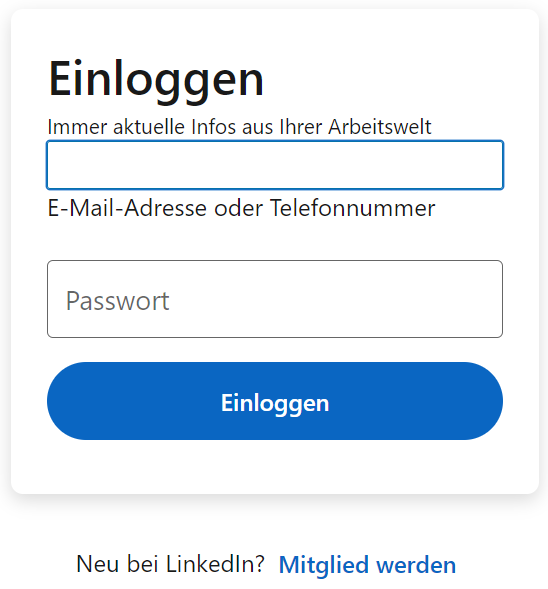}}  
\caption{Comparison of the page layout of the original site and the kit outcome.}
	\label{fig:images}
\end{figure*}

\subsection{Domain Name Registration}

Following the phishing kit deployment, we associated it with a domain name for enhanced legitimacy, using a cost-effective domain registrar that offers registration, management services, an API for domain purchases, and a Python library for seamless integration.

Initially, our attempts to register randomly generated domain names were thwarted by our registrar detection mechanisms.
As an alternative, we used a strategy involving an English dictionary, combining three words 
with hyphens, and adding a random number (e.g., \texttt{sun-car-blackhole-99.co}).
We refrained from typosquatting due to its potential detection by anti-phishing engines \cite{hook}.
Using an example domain registration code from the domain registrar quick start guide on PyPi, ChatGPT successfully created the necessary class.

During the registration, users must provide their registrant information. We have automated
the generation of random registration details using ChatGPT.

At first, we used the model to create identity attributes (name, phone, address, etc.) for testing purposes. Nevertheless, it consistently yielded the name ``John Doe," which could raise suspicion about the accuracy of the information.

We noticed that ChatGPT responded differently based on the language of the prompt. In 
another language, it exhibited better creativity and generated more authentic data. 
Therefore, we developed a Python function that introduced variability by randomizing 
country choices in the prompt.

As adding a domain name to our reverse proxy provider involves a change in the name servers,
we have sent our already developed Python class to ChatGPT and asked it to implement a method
to automatically change the name servers. 
We prompted ChatGPT to carefully analyze the transmitted code and it provided us with a clear and well-commented code. 

By default, the language model provides detailed explanations during code generation. 
Interestingly, when provided with a source code, ChatGPT optimizes it without considering a potential malicious intent.

\subsection{Reverse Proxy and CDN}

Integrating an online reverse proxy with our toolkit is interesting for several reasons. 
Our provider offers a free AntiBot service acting as a reverse proxy to conceal the real 
IP address of the (malicious) web server, while also providing a valid TLS certificate. We 
have used the TLS ``Flexible" mode for encryption between the client and the CDN, ensuring 
the green lock on browsers. 
While there is no encryption between the CDN and our web server, the reverse proxy handles 
the encryption, eliminating the need for a certificate on the web server. The provider 
Python library and API enable a comprehensive configuration.

To create the Python code for our toolkit, which automates the addition of a domain name 
to an online reverse proxy provider, we have supplied ChatGPT with the library 
documentation through a prompt.  
We have instructed the model to blend the provided documentation with its internal 
knowledge, highlighting that the documentation was up to date, unlike the knowledge base. 
Consequently, ChatGPT generated correct code.

\section{Results}

This section presents the results of testing the methodology.

\subsection{Testing the Generation of Phishing Sites \label{sec:tests}}

We first used ChatGPT to generate phishing sites 
for 80 popular brands \cite{comar}. The generation succeeded for 25 websites
(31.25\%) and failed for others. 
The number may appear as low, however, unsuccessful generation was 
due to the token size limits (16,384 tokens 
per prompt) in the GPT-3.5-turbo-16K model and the future plans of OpenAI include models 
with increased token limits that will allow handling larger websites.
Moreover, with potential code modifications and methodological adjustments (code 
segmentation), it is still feasible to create code that functions independently of the original source website code size.

The kit completes various steps within specific timeframes. On average, it takes 29 seconds to create a basic phishing page, involving cloning the original site, removing original sign-in buttons (e.g., Google Sign-In), cleaning forms, and integrating our Telegram API. The minification process, which eliminates unnecessary code parts and optimizes others, takes 57 seconds. Adding a modal requires 59 seconds while obfuscating an 800-character website content chunk takes an average of 14 seconds. Importantly, chunk obfuscation can be performed in parallel, maintaining consistent obfuscation times even for larger websites. As a result, the average total completion time for all these combined steps is approximately 4 minutes.
The outcomes demonstrate that ChatGPT exhibits consistent processing times regardless of task complexity. 

\subsection{Case Study: the LinkedIn Site \label{sec:li}}

As a case study, we have created a phishing website mimicking LinkedIn and assessed 
similarity across various pages.
During each stage of the generation process, we paused to convert the kit-generated pages 
into images and computed Structural Similarity (SSIM) scores between the output and the 
original website. SSIM is a metric used to gauge the resemblance between two 
images.
Figure~\ref{fig:images} presents the visual comparison between the outcomes of our kit at various stages and the original LinkedIn site.

The similarity stands at 100\% following the initial raw copy of the original site (see Figure~\ref{fig:images}a), 99.6\% after eliminating authentication elements and integrating our authentication API (see Figure~\ref{fig:images}b - adapted website), and 90.9\% after the modal inclusion (see Figure~\ref{fig:images}c). 
The disparity between copying and incorporating the modal is just 8.7\%, showcasing the balance between introducing deceptive modals while preserving visual fidelity.
ChatGPT obfuscation reduces SSIM to 87.6\% (see Figure~\ref{fig:images}d), with shifted 
text and missing CSS tags. As ChatGPT is not specifically trained for obfuscation, it 
occasionally omits some elements and struggles with distinguishing modifiable data. The 
segmented code also hampers context comprehension.

While the primary goal of the kit is to maintain visual similarity, the introduced 
modifications and obfuscation ideally contribute to lowering the detectability of 
malicious websites by code similarity-based anti-phishing engines. 
In this case, a lower code similarity value is more favorable (less chances to be 
detected).

For assessing source code similarity, we have used the Term Frequency-Inverse Document Frequency (TF-IDF) algorithm. Initially, the source code similarity between the original LinkedIn code and the adapted website was 93\%. However, following the obfuscation process using ChatGPT, the similarity significantly dropped to 56.6\%.

\subsection{Testing the Generation and Deployment\label{sec:tgd}}

Then, we have tested the generation of phishing websites as well as their 
deployment.
The process expects the OpenAI API key to interact with ChatGPT, a cloud provider API key, 
another one for the reverse proxy provider, an API key for the domain 
registration site, and a Telegram API key. 
During the experiments, we disabled the Telegram API to prevent the 
inadvertent collection 
of credentials from any visitor to the websites. The deployed websites used specific URLs, and we 
explained our experiment on the registered domain name website. We discussed the experiment with our domain registrar and obtained permission to conduct 
the study. We removed the phishing websites after deployment and tests. 

The entire process, involving generating phishing websites, crafting phishing kits, registering domain names, deploying the sites, and running the phishing emulation, averaged around 10 minutes for completion.

We have tested the complete generation and deployment process
of 20 instances of phishing websites on individually registered 
domain names. The deployment of 16 instances 
succeeded while 4 instances encountered issues during 
the phishing site generation (OpenAI errors like ``the models overloaded" or the creation 
of non-functional sites that could not be deployed).  

\section{Conclusions}

In this paper, we have shown that ChatGPT makes it possible to create and deploy a fully 
automated phishing kit by a person with little programming or development skills.
The requirement is to learn how phishing works and how to bypass the filters used by 
ChatGPT. 

To evaluate the efficacy of the generated phishing infrastructure, we have deployed 
several instances of phishing sites automatically in a matter of minutes.
Our case study demonstrated high quality of the generated phishing site, exhibiting a 
significant similarity to the original website.

A noteworthy constraint concerns the prevailing token limit that hampers the capability to 
generate and deploy phishing kits for larger websites. This issue is expected to be 
mitigated with future models accommodating 32k (and more) tokens or by adopting code 
segmentation strategies, requiring a deeper level of expertise.

The assessment shows that ChatGPT is not resilient to malicious usage despite 
extended safeguards and filters: 
adversaries can leverage ChatGPT to generate and deploy 
phishing websites swiftly, significantly increasing the potential risk associated with 
using ChatGPT for such illicit activities and expanding the reach and magnitude of phishing attacks. 

\balance

\bibliographystyle{IEEEtran}
\bibliography{main}

\end{document}